\documentstyle[index,twoside,epsf,12pt]{article}
\begin{document}

\begin{titlepage}
\begin{flushright}
UWThPh-1997-52 \\
27. Oktober 1997
\end{flushright}

\begin{center}

{\Large \bf History and outlook of statistical physics}
\footnote{Paper presented at the Conference on Creativity in
Physics Education, on August 23, 1997, in Sopron, Hungary.}
 \\[0.3cm]
{\bf Dieter Flamm} \\[3mm]
Institut f\"ur Theoretische Physik der Universit\"at Wien, \\
Boltzmanngasse 5, 1090 Vienna, Austria \\
Email: Flamm@Pap.UniVie.AC.AT \\[2cm]
{\large \bf Abstract}
\end{center}
\vspace*{6mm}
This paper gives a short review of the history of statistical physics
starting from D. Bernoulli's kinetic theory of gases in the 18th
century until the recent new developments in nonequilibrium kinetic
theory in the last decades of this century. The most important
contributions of the great physicists Clausius, Maxwell and Boltzmann
are sketched. It is shown how the reversibility and the recurrence paradox
are resolved within Boltzmann's statistical interpretation of the
second law of thermodynamics. An approach to classical and quantum
statistical mechanics is outlined. Finally the progress in
nonequilibrium kinetic theory in the second half of this century is
sketched starting from the work of N.N. Bogolyubov in 1946 up to
the progress made recently in understanding the diffusion processes
in dense fluids using computer simulations and analytical methods.

\end{titlepage}
In the 17th century the physical nature of the air surrounding the
earth was established. This was a necessary prerequisite for the
formulation of the gas laws. 
The invention of the mercuri barometer by Evangelista Torricelli
(1608--47) and the fact that Robert Boyle (1627--91) introduced
the pressure $P$ as a new physical variable where important steps.
Then Boyle--Mariotte's law
$P V$ = const. for constant temperature, where $V$ is the volume, was
formulated. \\
\begin{center}
{\large \sc First Kinetic Theory of Gases}
\end{center}
Daniel Bernoulli (1700--82) who had been born in Groningen in the
Netherlands and moved to Basle in Switzerland gave in 1738 in a
treatise on hydrodynamics a derivation of the gas laws from a
"`billiard ball"' model. He assumed that the gas consists of a very
large number of small particles in rapid motion. He already identified heat
with kinetic energy (living force). He then derived Boyle--Mariotte's
law for the gas--pressure at constant temperature on a movable piston
from the impact of the gas molecules onto the piston just in the same
way as it is still done today in elementary text books.
He also used the principle of conservation of mechanical energy and
concluded that, if the temperature changes, the pressure will also
change so that it is proportional to the square of the velocities of
the gas--particles and thus, for constant volume, will rise with rising
temperature. At that
time mechanical energy was still called vis viva in Latin or living
force in English. Bernoulli was, however, about a century ahead of his time
with his kinetic theory of gases. His model was almost forgotten. At
that time only his Swiss copatriots J.A. De Luc (1727--1817) and
George--Louis Le Sage (1724--1803) in Geneva and M.V. Lomonossov
(1711--65) in Russia mentioned it. As Stephen Brush points out in his book
on the kinetic theory ``The man who persuades the world
to adopt a new idea has accomplished as much as the man who conceived
that idea.''\footnote{S.G. Brush: Kinetic Theory, vol. 1. The Nature
of Gases and of Heat, Pergamon Press, London 1965, p. 9. Bernoulli's
paper and most other historic papers quoted below are reprinted in
vol. 1 or in vol.2: Irreversible Processes (1966) by S.G. Brush.}
The reason why Bernoulli's kinetic theory received so little attention was
that most scientists at that time believed in the so--called caloric
theory of heat.
\begin{center}
{\large \sc The Caloric Theory}
\end{center}
In the caloric theory heat was a substance called "'caloric"'.
Caloric was considered to be a fluid composed of particles which
repel each other. Like the earth is surrounded by its atmosphere each
matter particle was thought to be surrounded by an atmosphere of
caloric whose density increases with temperature. Thus at small
distances matter particles repel each other due to the repulsion of
their caloric atmospheres. In those days one did not know that matter
is held together by electrical forces but one thought that the
attractive forces between matter particles are of gravitational
origin just as the forces between the sun and the planets. At a
certain distance between matter particles there would be equilibrium
between the caloric repulsion and the gravitational attraction. As
the temperature rises more caloric is added to each matter particle
and consequently the caloric repulsion increases shifting the
equilibrium point outwards. In this way one could explain the thermal
expansion of matter including gases. It should be mentioned that such
reputed scientists as Pierre Simon, Marquis de Laplace (1749--1827)
gave a very sophisticated derivation of the gas laws within the
caloric theory.

At the end of the 18th century various arguments against the caloric
theory appeared. For instance: Does caloric have weight? In 1798
Benjamin Thompson (1753--1814) could show that the expected
additional weight when a body was heated could not be detected.
Another argument against the caloric theory was raised by Rumford
Humphry Davy (1778--1829). He remarked that an indefinite amount of
heat can be produced from matter by mechanical work, for instance by
friction. If caloric was a substance only a limited amount should be
available in matter. But these arguments did not really convince the
fans of the caloric theory.

Now let me mention the attempts for a revival of the kinetic theory. \\
\begin{center}
{\large \sc The Revival of the Kinetic Theory}
\end{center}
In 1820 in England
John Herapath, born 1790 at Bristol, critisized the derivation of the
gas laws by
Laplace and gave an account of the kinetic theory, but his paper was
not accepted for publication in the Philosophical Transactions of the
Royal Society because it was considered to be too speculative. It
finally appeared in 1821 in the Philosophical Transactions and would
have been forgotten if James Prescott Joule (1818--89) would not have
been influenced by it and by Herapath's book on Mathematical Physics
published in 1847.

In 1847 Joule published a paper in the Manchester Courier with the
title ``On Matter, Living
Force, and Heat'' where he stated the principle of conservation of
energy. From his experiments he concluded in this paper that heat is not a
substance but a form of energy: ``Experiment
has shown that whenever living force [kinetic energy] is apparently
destroyed or absorbed, heat is produced. The most frequent way in
which living force is thus converted into heat is by means of
friction.'' Furthermore he gave the amount of heat equivalent to the
converted kinetic energy. A year later he read another paper in
Manchester where he used Herapath's kinetic theory. It was not
published until 1851 and at first got very little attention until
Clausius quoted it in 1857.

The real breakthrough for the kinetic theory started in Germany after
Karl Kr\"onig (1822--79) published a paper about it in
Poggendorfs Annalen der Physik in 1856. Contrary to the widespread belief
that molecules of gas merely oscillate around definite positions of
equilibrium he assumed that they move with constant velocity in
straight lines until they strike against other molecules, or against
the surface of the container. Kr\"onig was actually a chemist but he
had a great reputation because he was editor of Fortschritte der
Physik, an annual review of physics, and he had great influence in
the German Physical Society. Kr\"onig's paper apparently motivated Rudolf
Clausius (1822--88) to publish on the kinetic theory. In fact Clausius 
writes that already before his first paper on heat in 1850 he had a
very similar conception of heat as Kr\"onig but in his former papers he
intentionally avoided mentioning this conception, because his
conclusions were deducible from general principles and did not depend
on these special conceptions. Clausius was
already well known from his papers on thermodynamics. In 1850 he had
given his verbal formulation of the second law that there exists no
thermodynamic transformation whose sole effect is to extract a
quantity of heat from a colder reservoir and to deliver it to a
hotter reservoir. In the
years until 1854 he had worked out its mathematical formulation. Since 1855 he
was professor at the ETH in Z\"urich. When Clausius started to work on the
kinetic theory it became fashionable. In 1857 he published his first
paper on the kinetic theory with the title
``The Nature of the Motion which we call Heat'' where he quoted the
papers of Kr\"onig and of Joule. The English translation of his paper
appeared in the same year in the Philosophical Magazine.
Two important arguments against the kinetic theory of heat were the
following: \\

How can heat traverse a vacuum if it is just irregular motion of
matter particles? There is no matter in the vacuum which could
propagate heat while the particles of caloric could easily penetrate
through the vacuum. \\

C.H.D. Buys--Ballot (1817--90) argued that since gas particles in the
kinetic theory move with velocities of a few hundred meters per
second one would expect that gases diffuse and mix much more rapidly
than observed. \\
  
In 1858 Clausius published a paper in which he could cope with the
second of these objections by introducing the mean free path of a
gas molecule. Gas molecules move at speeds of a few hundred meters an
hour but they undergo collisions with other gas molecules which
change their direction after a very short time of flight. The
actual distance they can move on the average freely along a straight
line in one direction is the mean free path $l$ given by 
\begin{equation}
l = \frac{3}{4} \frac{1}{n \pi {\sigma}^2} \qquad n = \frac{N}{V}
\end{equation}
where $n$ is the number density of gas molecules and $\sigma$ is the
diameter of the hard sphere particles which approximate the gas
molecules. For his estimate of the mean free path Clausius made the
drastic approximation that only one particle is moving and all others
are at rest. His result differs less than 10 per cent from the result
in Eq. (3) obtained by Maxwell one year later from a much
more refined derivation. Clausius who in
1865 introduced the concept of entropy continued to work on the
kinetic theory.

James Clerk Maxwell (1831--79), best known from his electromagnetic
field theory which he developed in the years from 1855 to 1873, read
his first paper on the kinetic theory in 1859 at a Meeting of the
British Association at Aberdeen. With the title ``Illustration of the
Dynamical Theory of Gases'' it appeared 1860 in print in the
Philosophical Magazine. While in earlier treatments the absolute
value of the velocities of the molecules was considered to be rather
uniform he was the first to assume a random motion for the molecules.
For thermal equilibrium he could then derive from symmetry
considerations his famous velocity distribution function which in
modern notation is given by 
\begin{equation}
f_0(\vec{v}) = n {\left(\frac{m}{2 \pi k T} \right)}^{\frac{3}{2}}
\exp{\frac{m {\vec v}^2}{2 k T}}
\end{equation}
where $\vec v$ is the velocity and $n$ the density of the molecules,
$m$ their mass, $k$ Boltzmann's constant and $T$ the absolute
temperature. For the mean free path he then obtained
\begin{equation}
l =  \frac{1}{\sqrt{2}} \frac{1}{n \pi {\sigma}^2}
\end{equation}
and for the viscosity of a dilute gas
\begin{equation}
{\eta}_0 (T) = \frac{1}{3} n m l \bar{v} \qquad with \qquad \bar{v} = {\left(
\frac{8 k T}{\pi m} \right)}^\frac{1}{2}
\end{equation}
where $\bar{v}$ is the mean absolute value of the velocity.
Inserting $l$ into the last equation he obtained a value independent
of the density and because of $\bar{v}$ proportional to the
square root of the absolute temperature
\begin{equation}
\eta_0 (T) = \frac{1}{3 \sqrt{2}} \frac{ m \bar{v}}{\pi {\sigma}^2} \;.
\end{equation}
The density independence of the viscosity was quite unexpected since
for a fluid the viscosity in general increases with increasing
density. After its experimental verification this result served as a strong
argument in favour of the kinetic theory.
\begin{center}
{\large \sc The Boltzmann Equation}
\end{center}
In 1872 Ludwig Boltzmann in Graz generalized Maxwell's approach for
the kinetic theory of dilute gases to non\-equilibrium processes,
so that he could investigate the transition from nonequilibrium to
equilibrium. His non--equilibrium single
particle distribution function $f \doteq f(\vec{x}, \vec{v}, t)$
gives the average number of molecules in a dilute gas at the position
$\vec{x}$ with velocity $\vec{v}$ at time $t$. The temporal change of this
distribution function consists of two terms, a drift term due to the
motion of the molecules and a collision term due to collisions with
other molecules. In the absence of an external field of force this
equation, which is now called Boltzmann equation, reads:
\begin{equation}
\frac{\partial f}{\partial t} = - \vec{v} \cdot \frac{\partial
f}{\partial \vec{x}} + J_B(ff) \; .
\end{equation}
Here $J_B(ff)$ is the binary collision term which takes only two
particle collisions into account, a good approximation for a dilute
gas. A further assumption in Boltzmann's
expression for the collision term is that the velocities of the
colliding molecules must be uncorrelated, which was later called the
assumption of ``molecular chaos'' by Jeans.

Now Boltzmann introduced the funtional 
\begin{equation}
H[f] = \int d^3 x \int d^3 v f(\vec{x}, \vec{v}, t) \log{ f(\vec{x},
\vec{v}, t)} 
\end{equation}
for which he could show
under very general assumptions for the intermolecular interaction that
if $f$ is a solution of Eq. (6) the time derivative of $H$ is always smaller
than zero or at most zero:
\begin{equation}
\frac{d H[f]}{d t} \le 0 \; .
\end{equation}
Furthermore for an ideal gas in equilibrium he could show that the
entropy $S$ is up to a sign proportional to $H$. For nonequilibrium
this is a generalization of the thermodynamic entropy now called
Boltzmann entropy
\begin{equation}
S(t) = - k H[f]
\end{equation}
and Eq. (8) is nothing but the second law of themodynamics for a
closed system
\begin{equation} 
\frac{d S(t)}{d t} \ge 0 \; .
\end{equation}
This is Boltzmann's famous $H$--theorem.

The $H$--theorem and the Boltzmann equation met with violent
objections from physicists and from mathematicians. These objections
can be formulated in the form of paradoxes. The most important ones
are the reversibility paradox formulated in 1876 by Boltzmann's
friend Josef Loschmidt (1821--95) and the recurrence paradox
formulated in 1896 by Ernst Zermelo (1871--1953). 
\begin{center}
{\large \sc The Reversibility Paradox}
\end{center}
In a paper in 1876 Loschmidt gave a recipe of how one can prepare an
initial condition
with decreasing entropy for any system which follows a motion with
increasing entropy.\footnote{J. Loschmidt, Sitzungsber. Kais. Akad.
Wiss. Wien, Math. Naturwiss. Classe {\bf 73}, 128--142 (1876).} One
need only reverse all its velocities at a
certain instant of time: $\vec{v}(t) \longrightarrow - \vec{v}(t)$.
This procedure is equivalent to time reversal and the argument is
usually called the reversibility paradox because the
equations of classical mechanics are invariant under time reversal
while the Boltzmann equation is not. The procedure of time reversal
violates the
hypothesis of ``molecular chaos'' since it is like a film which runs
backwards. All molecules which have just had a collision will collide
again and thus their velocities are correlated. This paradox
represented a severe objection
to the mechanical interpretation of the second law of
thermodynamics. Apparently there are just as many initial conditions
which lead at least for a short time to a decrease in the entropy of the
system as there are initial conditions leading to an increase
in the entropy. Why do
we never observe a decrease in entropy for large isolated systems? For
very small systems one can easily observe a decrease of entropy
in the form of statistical fluctuations, e.g. density
fluctuations, local pressure fluctuations or local temperature
fluctuations in very small regions of a gas. For very large systems
this is not the case because the small local fluctuations average
out. Statistical physicists have coined the term typicality for the
usual behaviour of macroscopic systems \footnote{J. L. Lebowitz:
Boltzmann's Entropy and Time's Arrow,
Phys. Today {\bf 46} (9), 32 (1993).}. If one pours, for
instance, a dye into a liquid it will gradually spread through the
whole liquid.
This behaviour is typical whenever you make such an experiment. It is
easy to tell the sequence in which snapshots of a
spreading dye were taken, even after their original order has been
deranged. How does this unidirectional behaviour in time come about?
For a very large
system, by far the largest number of states corresponds to equilibrium-
and quasi-equilibrium-states. The latter are states which differ
very little from the equilibrium state with maximum entropy and
cannot be distinguished macroscopically from the equilibrium state.
In our example of the liquid containing a dye they correspond to a
practically uniform distribution of the dye through the whole liquid
with very small local intensity fluctuations of the dye. With increasing size
of the system, the preponderance of the equilibrium- and
quasi-equilibrium-states becomes ever more overwhelming. If for every
possible state of a large system we put a marked sphere into an urn
and afterwards drew spheres from the urn indiscriminately, we would
practically always draw an equilibrium- or quasi-equilibrium-state.
The transition from nonequilibrium to equilibrium thus corresponds to
a transition from exceptionally rare nonequilibrium-states to
extremely probable states. This is Boltzmann's statistical
interpretation of the second law. 
\begin{center}
{\large \sc Statistical Mechanics}
\end{center}
Loschmidt's reversibility paradox led to a very fruitful discussion between
Boltzmann and Loschmidt about the second law and motivated Boltzmann
to work out his statistical interpretation of the second law in
detail.
To handle the reversibility paradox Boltzmann investigated
the entire phase space of a dynamical system consisting of $N$
particles or molecules. He found that the volume in the $6N$
dimensional phase space of the system which represents all possible
values of the three coordinates and the three momentum components of
each particle can be
subdivided into regions corresponding to macroscopic states of the
system which we shall call macrostates.

In his
paper of 1877 entitled ``On the relation between the second law of
the mechanical theory of heat and the probability calculus with
respect to the theorems on thermal equilibrium''\footnote{L.
Boltzmann: Sitzungsber. Kais. Akad. Wiss. Wien, Math. Naturwiss.
Classe {\bf 76} (1877) 373--435.}, Boltzmann now
presented a probabilistic expression for the entropy. He could show
that the entropy $S$ is proportional to the $6N$--dimensional phase space volume
$\Omega$ occupied by the corresponding macrostate of an $N$--particle
system: 
\begin{equation}
S \propto \log{\Omega}. 
\end{equation}
It is now usually written in the notation of Max Planck 
\begin{equation}
S = k \log W
\end{equation}
where $k$ is the Boltzmann constant and $W$ is
the number of microstates by which the macrostate of the system can be
realized.
This relation has been called Boltzmann's Principle by Albert
Einstein (1879--1955) in 1905 since it can be used as the foundation
of statistical mechanics. It is not limited to gases as Eq. (9) but
can also be applied to liquids and solid states. It can be obtained
from Eq. (11) by
introducing cells of finite volume in phase space as Boltzmann had
already done in order to obtain a denumerable set of microstates.
It implies that the entropy  is proportional to the logarithm of the
so--called thermodynamic probability $W$ of the macrostate which is
just the corresponding number of microstates.
A macrostate is determined by a rather small number of macroscopic
variables of the system such
as volume, pressure and temperature. The latter two correspond to
averages over microscopic variables of the system. A microstate, on
the other hand, is specified by the
coordinates and momenta of all molecules of the system. Due to the large
number of molecules there is a very large number of different choices
for the individual coordinates and momenta which lead to the same macrostate.
It turns out, that for a large system by far the largest number of
microstates corresponds to equilibrium-- and
quasi--equilibrium--states as we have already illustrated in the
example of the liquid containing a dye. The latter are states which differ
very little from the equilibrium state with maximum entropy and
cannot be distinguished macroscopically from the equilibrium state.
Thus this macrostate is the state of maximal entropy and the
transition from nonequilibrium to equilibrium corresponds to
a transition from exceptionally unprobable nonequilibrium-states to
the extremely probable equilibrium--state. In Boltzmann's statistical
interpretation the second law is thus not of absolute but only of
probabilistic nature. The appearance of so--called statistical
fluctuations in small subsystems was predicted by Boltzmann and he
recognized Brownian motion as such a phenomenon. The theory of
Brownian motion has been worked out independently by Albert Einstein
in 1905 and by Marian von Smoluchowski. The experimental verification of
these theoretical results by Jean Baptiste Perrin was important
evidence for the existence of molecules. \\

The term Statistical Mechanics has actually been coined by the great
American physicist J. Willard Gibbs (1839--1903) at a meeting of the American
Association for the Advancement of Science in
Philadelphia in 1884.\footnote{M.J. Klein: The Physics of J. Willard Gibbs in
his Time, Phys. Today, Sept. 1990, p. 40.} This was
one of the rare occasions when Gibbs went to a meeting
away from New Haven. He had been professor of mathematical physics at Yale
University since 1871 and had served nine years without salary. Only
in 1880, when he was on the verge of accepting a professoship at John
Hopkins University, did his institution offer him a salary.
He had realized that the papers of Maxwell and Boltzmann initiated a
new discipline which could be applied to bodies of arbitrary
complexity moving according to the laws of mechanics which were
investigated statistically. In the years following 1884 he formulated
a general framework for Statistical Mechanics and in 1902 published
his treatise.\footnote{J. W. Gibbs: Elementary Principles in
Statistical Mechanics. Developed with Especial Reference to the
Foundation of Thermodynamics. Yale Univ. Press 1902.}

Gibbs started his consideration with the  principle of conservation
of the phase space volume occupied by a statistical ensemble of mechanical
systems. He considered three types of ensembles.

The so--called microcanonical ensemble of Gibbs corresponds to an ensemble of
isolated systems which all have the same energy.
Boltzmann called this ensemble ``Ergoden''.\footnote{L. Boltzmann:
\"Uber die Eigenschaften monozyklischer und anderer damit verwandter
Systeme. Crelles Journal {\bf 98} (1884) p.68--94,
\"Uber die mechanischen Analogien des zweiten Hauptsatzes der
Thermodynamik. ibid
{\bf 100} (1887) p. 201--212 and Vorlesungen \"uber Gastheorie, II.
Teil, J.A. Barth, Leipzig 1898, p. 89.} In this case each member of
the ensemble coresponds to a different microstate and all microstates
have the same probability.

The canonical ensemble of Gibbs corresponds to systems in contact
with a heat bath. In this case the energy of the individual systems
is allowed to fluctuate around the mean value $E$. If $E_{\nu}$ is the
energy of an individual system $\nu$ of the ensemble, its probability
$P_{\nu}$ is proportional to an exponential function linear in the
energy $P_{\nu} \propto \exp{\left(- \frac{E_{\nu}}{k T}\right)}$
which is nowadays often called the Boltzmann factor.

For the grandcanonical ensemble of Gibbs not only the energy but also
the number of particles $N_{\nu}$ of the individual systems is
allowed to fluctuate around the mean value $N$.

If we introduce the density in $6N$--dimensional phase space for an
ensemble of physical $N$--particle systems 
\begin{equation}
\rho = \rho (\vec{x}_1 , \vec{p}_1 , \vec{x}_2 , \vec{p}_2 , ...
\vec{p}_N , \vec{x}_N , t)
\end{equation}
the Gibbs entropy can be written in the form
\begin{equation}
S = - k \int\limits_{V} d^3 x_1 d^3 x_2 ... \int\limits_{- \infty}^{\infty}
d^3 p_1 d^3 p_2 ... d^3 p_N \rho \log{\rho} .
\end{equation}
Introducing finite cells in phase space the number of microstates
becomes denumerable and will be labelled by $\nu = 1, 2, ... W$ where
$W$ is the total number of microstates. The expression for the
entropy then becomes
\begin{equation}
S = - k \sum_{\nu} P_{\nu} \log{P_{\nu}} \qquad with \qquad
\sum_{\nu} P_{\nu} = 1
\end{equation}
where $P_{\nu}$ is the probability of the corresponding microstate.
Eq. (15) has already the same form as the corresponding expression
for a quantum system with discrete energy levels. We may thus use
the procedure introduced by John von Neumann (1903--1957) in 1927 to determine
the equilibrium distribution $P_{\nu}$. It can be found by demanding that the
entropy Eq. (15) becomes a maximum under certain subsidiary
conditions which implies that the variation of $S$ with respect to
the $P_{\nu}$ vanishes.

For the microcanonical ensemble only the sum of all probabilities
must be one and if the total number of states is $W$ one obtains the
same probability  $P_{\nu} = \frac{1}{W}$ for all microstates which
implies that Eq. (15) reduces to Eq. (12). 

For the canonical ensemble the fluctuations of the energy $E_{\nu}$ of the
individual systems  around
the mean value $E$ requires the subsidiary condition
\begin{equation}
 \sum_{\nu} P_{\nu} E_{\nu} = E \ .
\end{equation}
The maximum entropy principle then gives for the probabilities
\begin{equation}
 P_{\nu}  = \frac{\exp{\left(- \frac{E_{\nu}}{k T}\right)}}{Z} \qquad
with \qquad Z = \sum_{\nu} \exp{\left(- \frac{E_{\nu}}{k T}\right)}
\end{equation}
where $T$ is the absolute temperature and $Z$ the canonical partition
function.

For the grandcanonical ensemble the fluctuations of
the number of particles $N_{\nu}$ of the individual systems
around the mean value $N$ require in addition to
Eq. (16) the subsidiary condition
\begin{equation}
 \sum_{\nu} P_{\nu} N_{\nu} = N \; ,
\end{equation}
which has to be multiplied by a Lagrange multiplier and added to the
entropy when the variation of $P_{\nu}$ is performed. This way one
obtains for the grand canonical probability distribution
\begin{equation}
 P_{\nu}  = \frac{\exp{\left( \frac{\mu N_{\nu} - E_{\nu}}{k T} \right)}}{\Xi} 
\qquad with \qquad \Xi
= \sum_{\nu} \exp{\left( \frac{\mu N_{\nu} - E_{\nu}}{k T}\right)}
\end{equation}
where $\mu$ is the chemical potential and $\Xi$ the grand canonical
partition function.

The quantum mechanical generalization of the framework of statistical
mechanics has already been given in 1927 by John von
Neumann who was born in Hungary. He introduced the density operator
$\hat{\rho} = \rho(\hat{p}_1 , \hat{x}_1 , ... \hat{p}_N , \hat{x}_N
, t)$ where the hat denotes the operator character. The quantum
mechanical generalization of Eq. (14) is then given by\footnote{J. v.
Neumann: G\"ottinger Nachr. (1927) 273.}
\begin{equation}
S = - k \mbox{Tr}(\hat{\rho} \log{\hat{\rho}})
\end{equation}
where Tr denotes the trace. If one uses for the trace eigenstates of the
Hamiltonian and the energy spectrum is discrete, the expression in Eq.
(20) reduces to the form given in Eq. (15). An important difference
between classical and quantum statistics is the symmetry of the state
under permutations of the particles. While for classical particles
permutations of the particles lead to a different state identical
quantum mechanical
particles can only be in an antisymmetric state if they are fermions
or in a symmetric state if they are bosons. An important application
of Eq. (20) is the treatment of ferromagnetism.

\begin{center}
{\large \sc Progress in Nonequilibrium Kinetic Theory}
\end{center}
Efforts for a systematic derivation as well as a generalization of
the Boltzmann equation have been made by N.N. Bogolyubov in
1946.\footnote{N.N. Bogolyubov: J. Phys. (USSR) {\bf 10} (1946) 256.}
The starting point for the derivation of the Boltzmann equation is
the time reversal invariant Liouville equation for the $N$--particle
phase space density in Eq. (13)
\begin{equation}
\frac{\partial \rho}{\partial t} = \{ H, \rho\}
\end{equation}
where $\{,\}$ is the Poisson bracket and $H$ the Hamiltonian of the
system. Eq. (21) follows from the classical equations of motion and
expresses the conservation of the probability in phase space. To arrive at the
Boltzmann equation which violates time reversal invariance, because
the direction of increasing entropy is singled out, some coarse
graining is necessary which is done by successively integrating over
the coordinates and momenta of $N - 1$ particles until one arrives at
the one particle distribution function $f \doteq f(\vec{x},
\vec{v}, t)$
which appeares in Eq. (6).\footnote{For a review see e.g. R.
Jancel: Foundations of Classical and Quantum Statistical Mechanics,
English translation ed. D. Ter Haar, Pergamon Press, Oxford 1969.}
This way one arrives at the so--called
B.B.G.K.Y. chain of equations which stands for the first letters of
the physicists Bogolyubov, Born and Green, Kirkwood, Yvon. Then one
has to perform the limit of low density, the so--called
Boltzmann--Grad--Limit and make the assumption of molecular chaos for
the initial distribution which implies factorization of the reduced
$n$--particle densities into products of one particle densities.
Furthermore one has to assume that the system is large enough so that
the influence of the walls of its container is negligible. There
exist a number of such derivations of the Boltzmann equation of which
I would like to mention the one by O.E. Lanford\footnote{ O.E.
Lanford: Time Evolution of Large Classical Systems, in
Dynamical Systems, Theory and Application, edited by J. Moser, Springer,
Berlin, 1975.} and one by an Italian group.\footnote{ R. Illner and M.
Pulvirenti: Global Validity of the Boltzmann
Equation for Two- and Three-Dimensional Rare Gas in Vacuum: Erratum and
Improved Result, Commun. Math.  Phys. {\bf 121} (1989) 143--146.} While
Lanford's derivation holds only for a very short time interval, the
Italian derivation implies so low densities that nearly no collisions
take place. These restrictions are not surprising since at higher
densities correlated collision sequences appear such as the ``ring
collisions'' shown in Fig. 1 which introduce
correlations between the colliding particles violating the assumption
of molecular chaos. In this way the relaxation to equilibrium is
slowed down. In computer simulations of a gas of hard spheres which
are often called ``relaxation experiments'' one can study the
approach to the one particle equilibrium distribution. Alder and
Wainwright did pioneering experiments of this kind in
1958.\footnote{B.J. Alder and T. Wainwright, in Transport Processes
in Statistical Mechanics, edited by I. Prigogine, Wiley--Interscience,
New York, 1958.} Quite recently such experiments have also been
performed at Vienna University.\footnote{Ch. Dellago and H.A. Posch:
Mixing, Lyapunov instability, and the approach to equilibrium in a
hard sphere gas, Phys. Rev. {\bf E 55} (1997) R9.} Following the
analytic methods of Bogolyubov a modified Boltzmann equation of the
type 
\begin{equation}
\frac{\partial f}{\partial t} = - \vec{v} \cdot \frac{\partial
f}{\partial \vec{x}} + J(ff) + K(fff) + L(ffff) + ...
\end{equation}
is obtained where $J(ff)$ containes the two particle collisions,
$K(fff)$ the three particle collisions, $L(ffff)$ the four paticle
collisions etc.. The solution of this generalized Boltzmann equation
with the Chapman--Enskog method then leads to a density or virial
expansion for the viscosity 
\begin{equation}
\eta(n,T) = \eta_0 (T) + n \eta_1 (T) + n^2 \eta_2 (T) + ...
\end{equation}
which is a power series expansion in the density $n$ with temperature
dependent coefficients. For higher densities, however, it turns out
that especially due to the ``ring collision'' terms the Bogolyubov
collision integrals $K(fff)$ and $L(ffff)$ become divergent and a cut
off for the mean free path has to be introduced. The revised density
expansion for the viscosity now contains a logarithmic term in the
density $n$:\footnote{For a review see e.g. E.G.D. Cohen: Kinetic Theory:
Understanding Nature through Collisions, in Thermodynamics and
Statistical Physics -- Teaching Modern Physics, ed. M.G. Velarde and
F. Cuadros, World Scientific, 1995.}
\begin{equation}
\eta(n,T) = \eta_0 (T) + n \eta_1 (T) + n^2 \ln{n} \, \eta'_2 (T) + n^2 \eta_2
(T) + ...
\end{equation}
Similar expressions result for the other transport coefficients.

\begin{center}
{\large \sc Dense Fluids}
\end{center}
With the development of more powerful computers extensive computer
simulations for dense hard sphere fluids have been performed in the
last twenty years.\footnote{W.W. Wood, in Fundamental Problems in
Statistical Mechanics, ed. E.G.D. Cohen, North--Holland Pub.,
Amsterdam 1975. } In this way, and from theoretical
considerations, it has been possible to identify two especially
relevant collision sequences going on in a dense hard fluid. These
are the ``cage diffusion'' collision sequence in Fig. 2 and the
``vortex diffusion'' collision sequence in Fig. 3. In ``cage
diffusion'' the particle finds itself trapped in a cage made up by
the surrounding particles and it requires several collisions with its
neighbours until it finds a hole to sneek out. It leads to a
significant change in the volume dependence of the viscosity of a
dense fluid as shown in Fig. 4\footnote{I.M. de Schepper, A.F.E.M.
Haffmans and J.J. van Loef: J. Stat. Phys. {\bf 57} (1989) 631.}.
Its importance was discussed
by de Schepper and Cohen in 1980.\footnote{I.M. de
Schepper and E.G.D. Cohen: Phys. Rev. {\bf A 22} (1980) 287; J. Stat.
Phys. {\bf 27} (1982) 223.} In ``vortex diffusion''
an energetic particle creates a vortex like a macroscopic sphere
moving through liquid, see Fig. 3. It was discovered around 1968 by
Alder and Wainwright\footnote{B.J. Alder and T.E. Wainright: Phys.
Rev. {\bf A1} (1979) 18.} and affects the long--time behaviour at
about twenty--fife mean free times. The above mentioned collision sequences
constitute corrections to the Boltzmann equation in dense fluids and
provide an understanding of the behaviour of dense fluids through collisions.
\begin{center}
{\large \sc Acknowledgement}
\end{center}
The author would like to thank Prof. H.A. Posch for critical
reading of the manuscript.
\begin{figure}
\begin{picture}(14.5,7.5)
\put(-0.2,0){
\epsfysize=7.5cm
\epsffile{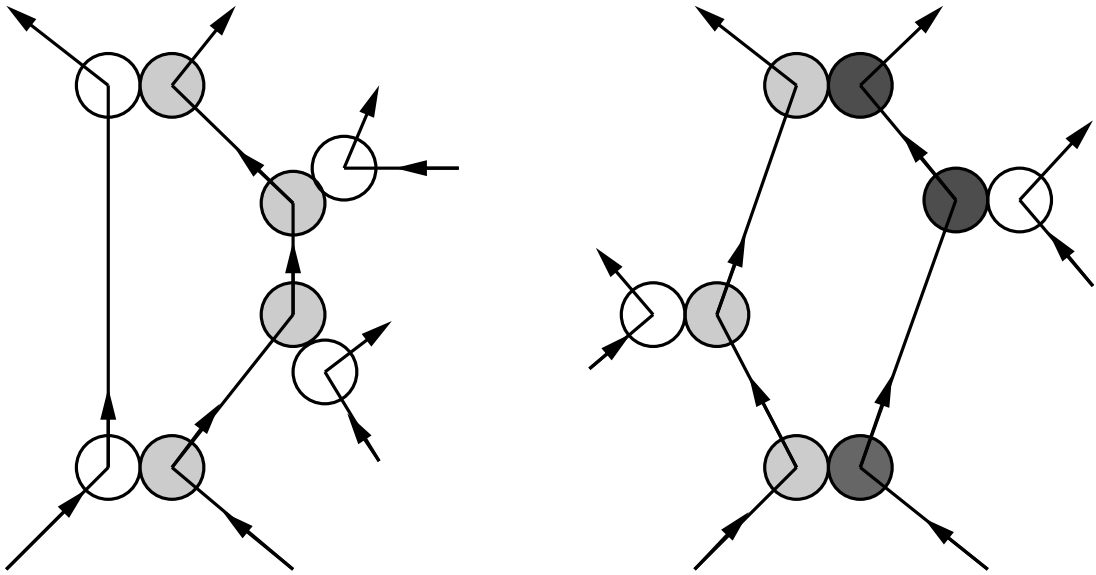}}
\end{picture}
\caption{Examples of correlated four particle ``ring collision''
sequences.}
\end{figure}

\begin{figure}[h]
\begin{picture}(14.5,8)
\put(3,0){
\epsfysize=8cm
\epsffile{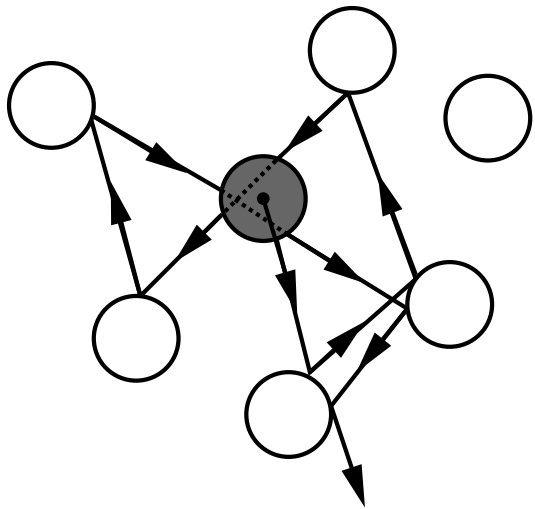}}
\end{picture}
\caption{Example of a correlated ``cage diffusion'' collision sequence.}
\end{figure}

\begin{figure}
\begin{picture}(14.5,9)
\put(-0.6,0){
\epsfysize=8cm
\epsffile{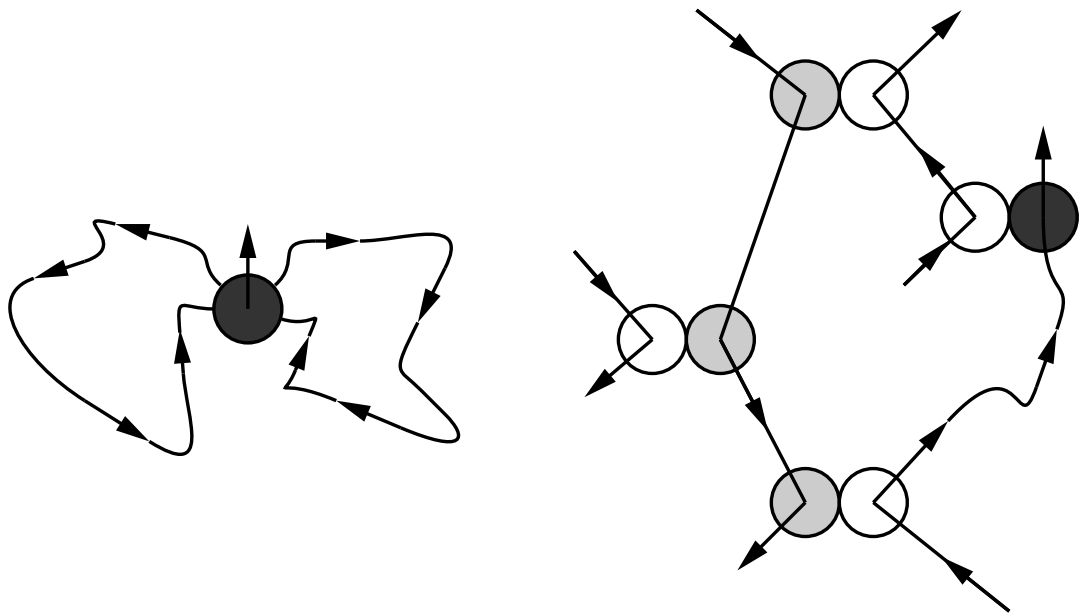}}
\end{picture}
\caption{Example of a correlated ``vortex diffusion'' collision
sequence. (a) Two vortex rings. (b) Left vortex ring in more detail.}
\end{figure}

\begin{figure}
\begin{picture}(14.5,9.5)
\put(1.5,0){
\epsfysize=9.5cm
\epsffile{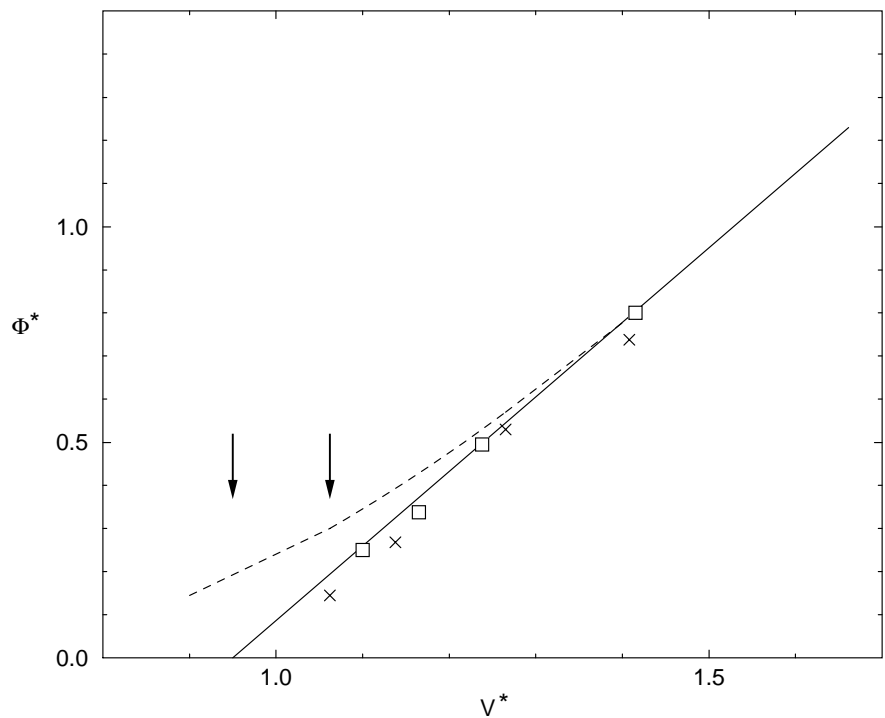}}
\end{picture}

\caption{The reduced hard--sphere fluidity $\phi ^* =
\frac{\eta_0}{\eta}$, which corresponds to the inverse of the viscosity
$\eta$ multiplied with the low density limit $\eta_0$ of the
viscosity, is plotted as a function of the
reduced volume $V^*$ which is the volume of the liquid divided by the
volume of the hard spheres. The dashed curve is $\phi ^*$ according to
the classical Enskog theory  and the solid curve according to the
mode coupling theory which takes also ``cage diffusion'' into
account. The squares and crosses are from computer simulation experiments.}
\end{figure}
\end{document}